\documentclass{article}
\usepackage{spconf,amsmath,amssymb,graphicx,hyperref}
\usepackage[nolist,nohyperlinks,printonlyreused]{acronym}
\usepackage{tcolorbox} 
\usepackage{float}
\usepackage{subcaption}
\title{Task-Based Adaptive Transmit Beamforming for Efficient Ultrasound Quantification}
\name{Oisín Nolan, Wessel L. van Nierop, Louis D. van Harten, Tristan S.W. Stevens, Ruud J.G. van Sloun\thanks{This work was supported by the European Research Council (ERC) under the ERC starting grant nr. 101077368 (US-ACT).}}
\address{Dept. of Electrical Engineering, Eindhoven University of Technology, The Netherlands}
\begin{document}
%\ninept
%
\maketitle
\begin{abstract}
Wireless and wearable ultrasound devices promise to enable continuous ultrasound monitoring, but power consumption and data throughput remain critical challenges. Reducing the number of transmit events per second directly impacts both. We propose a task-based adaptive transmit beamforming method, formulated as a Bayesian active perception problem, that adaptively chooses where to scan in order to gain information about downstream quantitative measurements, avoiding redundant transmit events. Our proposed Task-Based Information Gain (TBIG) strategy applies to any differentiable downstream task function. When applied to recovering ventricular dimensions from echocardiograms, TBIG recovers accurate results using fewer than 2\% of scan lines typically used, showing potential for large reductions in the power usage and data rates necessary for monitoring. Code is available at \url{https://github.com/tue-bmd/task-based-ulsa}.
\end{abstract}
\begin{acronym}
\acro{TBIG}{Task-Based Information Gain}
\acro{GIG}{General Information Gain}
\acro{MSE}{Mean Squared Error}
\acro{LVID}{Left Ventricular Inner Diameter}
\acro{LVPW}{Left Ventricular Posterior Wall}
\acro{IVS}{Interventricular Septum}
\acro{DPS}{Diffusion Posterior Sampling}
\acro{PLAX}{parasternal long-axis}
\acro{MAE}{mean absolute error}
\end{acronym}
% Variables
\newcommand{\T}{T} % sequence length
\newcommand{\action}{a}
\newcommand{\act}{\action}

\newcommand{\y}{\mathbf{y}} % observation
\newcommand{\z}{\mathbf{z}} % latent
\newcommand{\w}{w}
\newcommand{\uw}{\widetilde{\w}}
\newcommand{\parti}{i} % particle index
\newcommand{\ti}{t} % time index

% bold monospaced font
\newcommand{\code}[1]{\textbf{\texttt{#1}}}

% shorthand
\newcommand{\taui}{\tau_{\text{init}}}

% vector-valued variables
\newcommand{\ba}{\mathbf{a}}
\newcommand{\bd}{\mathbf{d}}
\newcommand{\bh}{\mathbf{h}}
\newcommand{\bk}{\mathbf{k}}
\newcommand{\bn}{\mathbf{n}}
\newcommand{\bx}{\mathbf{x}}
\newcommand{\by}{\mathbf{y}}
\newcommand{\bz}{\mathbf{z}}
\newcommand{\bep}{\mathbf{\epsilon}}

% sans serif
\newcommand{\sfi}{\mathsf{i}}

% tensor variables
\newcommand{\tens}[1]{\bm{\mathsf{#1}}}
\newcommand{\tA}{\tens{A}}
\newcommand{\tE}{\tens{E}}
\newcommand{\tY}{\tens{Y}}
\newcommand{\tX}{\tens{X}}

% Variables ++
\newcommand{\allActions}{\fullSeq{\action}}

% Common arguments
\newcommand{\fullSeq}[1]{{#1}_{1:\T}}

% Functions
\newcommand{\entropy}{H}

\newcommand{\proposal}{q\left(\particle~\mid~\particle[t-1], \mathbf{y}_t\right)}
\newcommand{\lproposal}{q_\theta \left(\particle~\mid~\particle[t-1], \mathbf{y}_t\right)} % learned proposal
\newcommand{\ltransition}{p_\phi\left(\particle~\mid~\particle[t-1]\right)}

\newcommand{\likelihood}{p\left(\mathbf{y}_t~\mid~\particle\right)}
\newcommand{\transition}[1][i]{p\left(\state^{#1}~\mid~\state[t-1]^{#1}\right)}

\newcommand{\state}[1][\ti]{\z_{#1}}
\newcommand{\particle}[1][\ti]{\z_{#1}^{i}}
\newcommand{\particleWeight}[1][\ti]{\w_{#1}^\parti}

\newcommand{\elbo}{\hat{\ell}(\theta)}
\newcommand{\posterior}{p(\z_t~\mid~\y_{1:t})}

% General
\newcommand{\real}{\mathbb{R}}

% circled numbers
\newcommand{\circled}[1]{%
  \tcbox[
    colback=white,                 % White background for the circle
    colframe=black,                % Black frame (the edge)
    boxrule=0.4pt,                 % Thickness of the frame
    arc=5pt,                     % <--- KEY: Set 'arc' to a very large value to force full rounding
    outer arc=5pt,               % Ensures the outer edge also rounds fully
    boxsep=0pt,                    % No default internal padding
    left=2.5pt, right=2.5pt,       % <--- KEY: Balance horizontal padding (adjust as needed)
    top=1.5pt, bottom=1.5pt,       % <--- KEY: Balance vertical padding (adjust as needed)
    nobeforeafter,                 % Essential for inline use
    halign=center,                 % Center content horizontally
    valign=center,                 % Center content vertically
    % No 'square=true' needed here, as the arc and balanced padding will create the circle
  ]{\textsf{\textbf{#1}}}                 % The content: a bold version of your number
}
\section{Introduction}
Ultrasound is a popular modality for medical imaging, offering high temporal resolution, cost-effectiveness, and versatility \cite{chan2010basics}. Another notable advantage is portability, with recent research showing promising results for wireless and wearable ultrasound technology \cite{garcia2025survey, huang2023emerging}, for example, a wearable ultrasound patch used for continuous monitoring at the ICU. The past years have shown great development in ultrasound patch hardware \cite{lin2024fully}, including electronics miniaturization and skin adhesive materials. Yet, power consumption and data communication bottlenecks remain important challenges for long-term monitoring \cite{huang2023emerging}. These, in turn, are strongly dependent on the number of transmit events per second (pulse repetition interval), with more events typically yielding higher image quality at the expense of the aforementioned energy and communication bottlenecks. Reducing these while retaining performance can be achieved via smart acquisition strategies and powerful reconstruction algorithms \cite{van2025patient}. We note further that while a typical ultrasound exam involves acquiring images of the anatomy, the aim in continuous monitoring is to track some quantitative parameters of the anatomy. In this paper, we leverage this fact to develop an algorithm that adaptively steers the ultrasound transmit beam to image only parts of the anatomy that contain information about a quantitative parameter of interest, $D_t$. In this way, we directly reduce the required number of transmit events, thereby reducing power usage and data rates, while acquiring all relevant information for accurate estimation of $D_t$. 

We model this as a Bayesian \textit{active perception} problem \cite{van2024active}, wherein the ultrasound probe becomes a sensing agent that tracks a probability distribution representing its beliefs about the state of $D_t$ over time. It is active in the sense that it \textit{chooses} which scan lines to acquire in order to minimize its own uncertainty about $D_t$. The agent does this in an iterative fashion, alternating between action and perception in what is referred to as a \textit{perception-action loop}. We refer to this process as \textit{task-based adaptive transmit beamforming}, given that transmit pattern is optimized with respect to some downstream measurement task, with the sole criterion for the measurement task being that it is a differentiable function of the ultrasound image. Our contributions are summarized as follows:
\begin{itemize}
    \item We derive and implement a novel perception-action loop that drives generic task-based transmit beamforming strategies to recover quantitative parameters with minimal transmit events.
    \item We evaluate our model using measurements quantifying ventricular hypertrophy produced by the EchoNet\-LVH model \cite{duffy2022high}, showing that our algorithm can recover accurate estimates of the measurement signal using a small fraction ($\sim$2\%) of the scan lines typically used for imaging.
\end{itemize}
% An example of such a device is a wearable ultrasound patch, used for continuous monitoring outside the hospital.
\section{Related Work}
\begin{figure*}[!t]

    \centering
    \includegraphics[width=\linewidth]{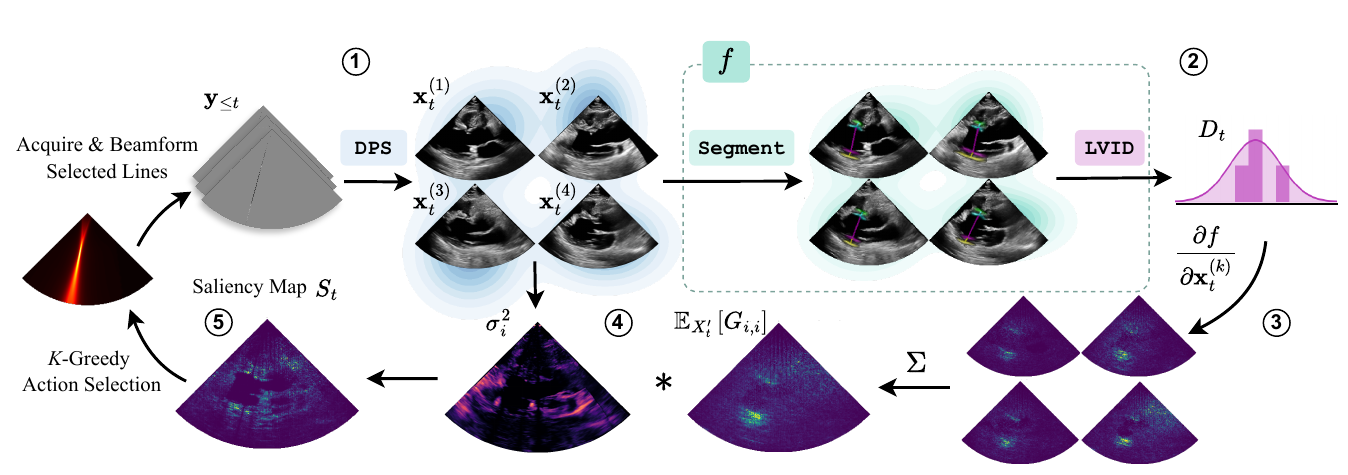}
    \caption{Diagram illustrating single iteration of the task-based perception-action loop using EchoNetLVH segmentation for the downstream task. \circled{1} Generate a set of posterior samples from the sparse acquisition using \ac{DPS}. \circled{2} Pass each posterior sample $\bx_t^{(i)}$ through the downstream task model $f$ to produce samples from the downstream task distribution. \circled{3} Compute the Jacobian matrix using each of the posterior samples as inputs. \circled{4} Average those Jacobian matrices and multiply them with the pixel-wise variance of the input images to produce the downstream task saliency map. \circled{5} Apply $K$-Greedy Minimization to select $K$ scan line locations for the next acquisition.}
    \label{fig:diagram}
\end{figure*}

A related research direction uses compressed sensing to reduce the data rates required for ultrasound imaging \cite{yousufi2019application}. A number of these works use supervised machine learning methods to recover beamformed images from subsampled channel data using fixed or random subsampling strategies that are independent of the data, learning only the reconstruction model \cite{mamistvalov2022deep, khan2020adaptive}. Huijben \textit{et al.} \cite{huijben2020learning} propose a method for learning both the sampling and reconstruction models in an end-to-end manner, where the reconstruction model can be some downstream function of the fully-sampled image. Due to the supervised nature of these methods, however, they need to be trained on datasets consisting of the target subsampling rates and downstream tasks, effectively requiring a re-training of any learned downstream task model, and making them susceptible to out-of-distribution errors if the subsampling rate or task change at test time. Furthermore, the black-box nature of both the sampling policy limits the explainability of the resulting subsampling strategies.
% Van Gorp \textit{et al.} \cite{van2021active} generalize the concept to an active setting, though their approach is not demonstrated on ultrasound.
An alternative approach is that of \textit{cognitive ultrasound} \cite{van2024active}, where acquisition is driven by active perception In this paradigm, the acquisition process is formulated as a perception–action loop, in which beamforming or sampling decisions are chosen to minimize uncertainty about a latent parameter of interest, typically using an explicit observation model for perception, and a white-box action selection function, enhancing interpretability. Federici \textit{et al.} \cite{federici2024active} exemplify this approach by designing an active perception algorithm for adaptive beam steering to track the fetal heart and estimate heart rate from power Doppler observations. While the objective is similar to ours -- recovering a downstream measurement -- their observation model is tailored to Doppler acquisitions and is not applicable to tasks defined on B-mode images. Van Nierop \textit{et al.} \cite{van2025patient} propose a cognitive beamforming algorithm that adaptively selects transmit patterns to minimize uncertainty about the reconstructed B-mode image, leveraging state of the art generative image models to implement perception on images. In this work, we build upon the cognitive ultrasound framework introduced by van Nierop \textit{et al.}, but move beyond image fidelity as the sole target, generalizing the approach to downstream quantification functions defined on B-mode images. In our experiments, we benchmark against their method, dubbed \acf{GIG}.
\section{Method}
\label{sec:method}
In this section, we derive a novel perception-action loop for task-based transmit beamforming, where the goal is ultimately to minimize uncertainty about $D_t$. We first introduce a way to quantify that uncertainty, which we show can be decomposed into a sum over a saliency map defined in the image domain, identifying the degree to which each pixel in the image space $X_t$ contributes to the uncertainty in $D_t$, thereby indicating tissue locations that should be targeted in the next transmit event. Fig. \ref{fig:diagram} provides a visual overview.

\subsection{Perception}
The goal of the perception step is to infer a probability distribution over possible values for $D_t$ given the scan lines acquired so far. Our perception step closely follows that of van Nierop \textit{et al.} \cite{van2025patient}, implementing Bayesian inference via the \acf{DPS} algorithm \cite{ChungKMKY23}, which uses a diffusion model \cite{ho2020denoising} to generate samples from the posterior distribution over fully-observed images given partial observations of recent frames $\bx_t \sim p(X_t \mid \by_{\leq t})$. Once a set of posterior samples $\{\bx_t^{(i)}\}_{i=1}^{N_p}$ has been generated at time $t$, the downstream task distribution $D_t \mid \by_{\leq t}$ is approximated by a set of samples $\{\bd_t^{(i)}~=~f(\bx_t^{(i)})\}_{i=1}^{N_p}$ produced by passing each posterior image sample through the downstream task model $f$.

\subsection{Action}
In order to drive action selection, we first quantify the uncertainty $U(D_t)$ in the state of $D_t$ using the scalar quantity $\mathbb{E}\left[||D_t - \mathbb{E}[D_t]||_2^2\right] = \text{tr}(C^D_t)$ measuring the expected Euclidean distance of the downstream task variable $D_t$ from its mean, with $C^D_t$ being the covariance of $D_t | \by_{\leq t}$. %Then, by modeling the covariance matrix for the posterior distribution $X_t \mid \by_{<t}$ as diagonal (i.e. pixel-wise), 
Then, by approximating the downstream task function $f$ with a first-order Taylor expansion, we have that 
%we have that $C^D_t$, can be computed as a linear function of the covariance of $X_t \mid \by_{<t}$, denoted $C^X$. In particular, we approximate 
$C^D \approx J C^X J^\top$, where $J$ is the Jacobian matrix of $f$ with reference point $X'_t~\sim~p(X_t~\mid~\by_{\leq t})$, $X_t$ and $X'_t$ are independent random variables sampled from the posterior, and $C^X_t$ is the posterior covariance of the images $X_t \mid \by_{\leq t}$. We use a random variable as the reference point to ensure that the Taylor expansion is computed at valid input points, rather than, for example, the posterior mean $\mathbb{E}\left[X_t \mid \by_{\leq t}\right]$, which may be off-manifold and out-of-distribution for the downstream task model. If we approximate $C^X_t$ as diagonal, we can then show that $U(D_t)$ can be computed as a sum of saliency-weighted variances from the pixels in the input image $X_t$:
\begin{align}
U(D_t)
  &= \mathbb{E} \left[||D_t - \mathbb{E}[D_t]||_2^2\right] \nonumber \\
  &= \text{tr}(C^D) 
     & & \nonumber \\
  &\approx \text{tr}(JC^X J^\top) 
     & & \nonumber \\
  &= \text{tr}(J^\top J C^X) 
     & & \text{(cyclic property of trace)} \nonumber \\
  &= \text{tr}(G C^X) 
     & & \text{($G := J^\top J$)} \nonumber \\
  &= \sum_i G_{i,i} \sigma^2_i
     & & \text{($\sigma^2_i := C^X_{i,i}$)}
\end{align}
This decomposition therefore becomes a sum of pixel-wise variances in $X_t$, denoted $\sigma^2_i$, weighted by diagonal entries of the Gram matrix $G_{i, i}$. Recall now that the Jacobian matrix is a function of the random variable $X'_t$. This means that $U(D_t)$, is also a random variable that is a function of $X'_t$. In order to get a scalar objective to optimize, we therefore take the expected value of $U(D_t)$, denoted $\bar{U}(D_t)$, employing the law of the unconscious statistician (LOTUS) \cite{pishro2014introduction}:
\begin{align}
\label{eq:concrete}
\bar{U}(D_t) &= \mathbb{E}_{U(D_t)}\left[ U(D_t) \right]  \nonumber \\
&= \mathbb{E}_{X'_t}\left[ U(D_t) \right]   & & \text{(LOTUS)} \nonumber \\
  &= \mathbb{E}_{X'_t}\left[\text{tr}(G C^X)\right] \nonumber \\
&= \text{tr}(\mathbb{E}_{X'_t}[G] C^X) & & \text{(\cite{soch2024statproofbook})} \nonumber \\
&= \sum_i \mathbb{E}_{X'_t}[G_{i,i}] \sigma^2_i \nonumber \\
&\approx \frac{1}{N} \sum_i \sum_jG_{i,i} \big|_{\bx = \bx^{(j)}} \sigma^2_i
     & & \text{(Monte Carlo)}
\end{align}
In other words, we average the Jacobian matrices computed with a set of reference points $\{\bx^{(j)}\}_{j=1}^{N_p}$. In practice, we can also estimate $\sigma^2_i$ as the empirical variance from the same set of samples. The diagonal of $G$ then contains the L2-norms of the columns of this averaged Jacobian, intuitively quantifying the overall impact that the $i^{th}$ input pixel in $X_t$ has in determining the value of $D_t$.

Our goal is then to choose a set of scan line locations $A_t$ for the next transmit event that will minimize $\bar{U}(D_t)$. The focusing angles affect $\bar{U}(D_t)$ via $\sigma_i^2$, specifically in that observing the $i^{th}$ pixel eliminates its variance, setting $\sigma_i^2 = 0$. Along with this effect, the generative perception model may resolve uncertainty in other pixels whose values can be inferred from the new observation, and certain other pixel variances may increase due to unpredictable temporal dynamics. Accounting for the entire impact of all candidate measurement actions on $\bar{U}(D_t)$ is therefore computationally challenging, potentially requiring the agent to simulate its perception step for the measurements resulting from each possible choice of $A_t$. Instead, we opt for an approximate greedy algorithm, \textit{$K$-Greedy Minimization}, introduced by van Nierop \textit{et al.} \cite{van2025patient}, using the saliency map $S_t = [\mathbb{E}_{X'_t}[G_{1,1}] \sigma^2_1, \dots, \mathbb{E}_{X'_t}[G_{i,i}] \sigma^2_i, \dots, \mathbb{E}_{X'_t}[G_{N,N}] \sigma^2_N]$ as the input, rather than a pixel-wise entropy map. The application of K-Greedy Minimization then returns a set of $K$ scan line locations to be transmitted at time $t+1$, completing the action step.
\section{Results}
\begin{figure*}[!t]
    \centering
    % Subfigure (a)
    \begin{subfigure}[t]{0.66\linewidth}
        \centering
        \includegraphics[width=\linewidth]{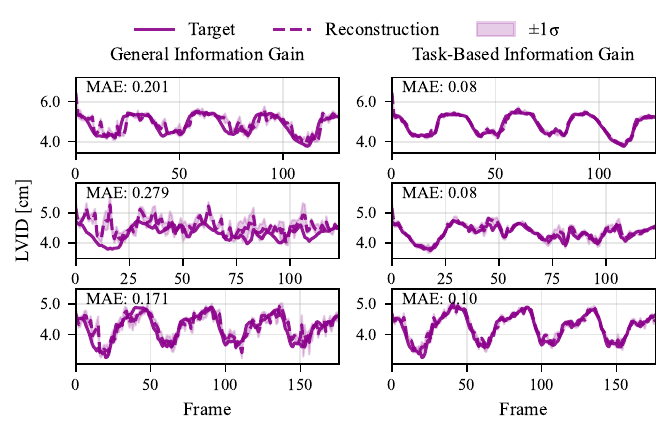}
        \caption{}
        \label{fig:time-series-a}
    \end{subfigure}%
    \hfill
    % Subfigure (b)
    \begin{subfigure}[t]{0.32\linewidth}
        \centering
        \includegraphics[width=\linewidth]{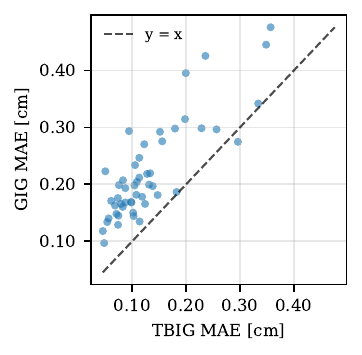}
        \caption{}
        \label{fig:time-series-b}
    \end{subfigure}
    \caption{(a) shows a qualitative comparison of measurement signal recovery for three patients using (i) \acs{TBIG} and (ii) \ac{GIG} sampling strategies with 5/256 scan lines. The \ac{MAE} between the target and reconstruction is provided at the top left. The uncertainty for each reconstruction is quantified as the standard deviation of the measurement values estimated from the samples in the belief set computed by applying $f$ to $\{\bx_t^{(i)}\}_{i=0}^{N_p}$ for both methods. (b) shows a sample-wise error comparison for each patient in the evaluation set, where it is clear that \acs{TBIG} almost always outperforms \ac{GIG}.}
    \label{fig:time-series}
\end{figure*}

\begin{figure}[!t]
    \centering
    \includegraphics[width=\linewidth]{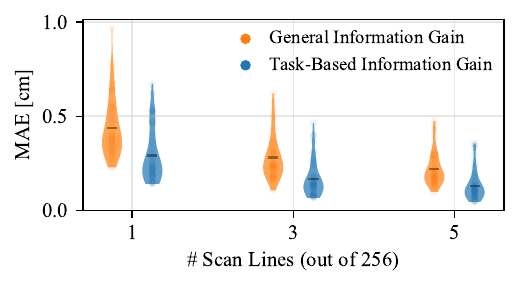}
    \caption{The distribution of \ac{MAE} scores between target and reconstructed \ac{LVID} time series for the first 100 frames from 50 patients from the EchoNetLVH validation set for both \acs{TBIG} and \acs{GIG} strategies.}
    \label{fig:measurement-mae}
\end{figure}
In order to evaluate the performance of our algorithm in a practical setting, we apply it to the task of recovering measurements of heart dimensions typically taken from the \ac{PLAX} view in echocardiography; in particular, the \acf{LVID}. We use the EchoNetLVH model \cite{duffy2022high} to produce heatmaps for the location of each of the measurement anchor points, with the center of mass of each heatmap identifying its anchor point, and the Euclidean distance between anchor points producing the eventual measurement. The measurement process producing $\by_{\leq t}$ is simulated using a masking operation, and a diffusion model trained over sequences of 3 frames from the EchoNetLVH dataset is used for the perception step, following the architecture of van Nierop \textit{et al.} \cite{van2025patient}. This was implemented using the \texttt{zea} python package \cite{zea2025}.

We compare our perception-action loop introduced in Section \ref{sec:method}, \acf{TBIG}, to the perception-action loop introduced by van Nierop \textit{et al.} \cite{van2025patient}, \acf{GIG}. The perception model in both methods uses a SeqDiff \cite{stevens2025sequential} initialization step of $\tau_{\texttt{SeqDiff}}=50$, with a total of 500 diffusion steps. The aim of the experiments is to evaluate whether explicitly optimizing the selected transmit angles for a downstream task results in a significantly improved estimate of the downstream parameters of interest. We compare the two strategies qualitatively and quantitatively in terms of the \ac{MAE} between the target signal, acquired by applying $f$ to the fully-sampled B-mode images, and the reconstructed signal, acquired by applying the EchoNetLVH model to $\{\bx_t^{(i)}\}_{i=1}^{N_p}$, averaging the resulting $N_p$ heatmaps for each anchor point, and extracting the measurements as described above.

In Fig. \ref{fig:time-series} we compare the downstream measurement signals recovered by \ac{TBIG} and \ac{GIG} to the target signals, using a budget of 5/256 scan lines. It is clear that the \ac{TBIG} objective recovers a significantly more accurate estimate of the target signal in almost all cases. In Fig. \ref{fig:measurement-mae}, we report the distributions of the \ac{MAE} evaluated on sequences of 100 frames from a set of 50 unseen patients from the EchoNetLVH validation set, using a number of different subsampling rates (1, 3, and 5 lines out of a total of 256).
\section{Discussion \& Conclusion}
It is clear from the results that \acf{TBIG} outperforms \acf{GIG} across all measurement targets and subsampling rates. We observe also that \ac{TBIG} using only 3 lines outperforms \ac{GIG} using 5, indicating the potential for a significant reduction in data rate relative to \ac{GIG} without sacrificing quality. We also highlight that the measurement signals have been recovered here using only a tiny fraction of the scan lines typically used (less than 2\%), indicating potential for use in continuous monitoring and other wireless ultrasound technologies. A promising avenue for future work is therefore to apply this method in a continuous monitoring setting, using data from a wearable ultrasound patch, and evaluate its effect on battery life and data throughput.


\clearpage
\bibliographystyle{IEEEbib}
\bibliography{strings,refs}

@book{pishro2014introduction,
  title={Introduction to probability, statistics, and random processes},
  author={Pishro-Nik, Hossein},
  year={2014},
  publisher={Kappa Research, LLC Blue Bell, PA, USA}
}

@misc{soch2024statproofbook,
  author       = {Soch, Joram and others},
  title        = {StatProofBook/StatProofBook.github.io: The Book of Statistical Proofs},
  year         = {2024},
  howpublished = {\url{https://statproofbook.github.io/}},
  note         = {Version 2023, Zenodo: \url{https://doi.org/10.5281/zenodo.4305949}}
}

@article{van2025patient,
  title={Patient-Adaptive Focused Transmit Beamforming using Cognitive Ultrasound},
  author={van Nierop, Wessel L and Nolan, Ois{\'\i}n and Stevens, Tristan SW and van Sloun, Ruud JG},
  journal={arXiv preprint arXiv:2508.08782},
  year={2025}
}

@article{garcia2025survey,
  title={Survey on wireless ultrasound imaging},
  author={Garc{\'\i}a, Laura and Viciano-Tudela, Sandra and Lloret, Jaime and Molt{\'o}-Jorda, Jos{\'e}-Manuel},
  journal={Health and Technology},
  pages={1--12},
  year={2025},
  publisher={Springer}
}

@article{huang2023emerging,
  title={Emerging wearable ultrasound technology},
  author={Huang, Hao and Wu, Ray S and Lin, Muyang and Xu, Sheng},
  journal={IEEE Transactions on Ultrasonics, Ferroelectrics, and Frequency Control},
  volume={71},
  number={7},
  pages={713--729},
  year={2023},
  publisher={IEEE}
}

@incollection{chan2010basics,
  title={Basics of ultrasound imaging},
  author={Chan, Vincent and Perlas, Anahi},
  booktitle={Atlas of ultrasound-guided procedures in interventional pain management},
  pages={13--19},
  year={2010},
  publisher={Springer}
}

@article{lin2024fully,
  title={A fully integrated wearable ultrasound system to monitor deep tissues in moving subjects},
  author={Lin, Muyang and Zhang, Ziyang and Gao, Xiaoxiang and Bian, Yizhou and Wu, Ray S and Park, Geonho and Lou, Zhiyuan and Zhang, Zhuorui and Xu, Xiangchen and Chen, Xiangjun and others},
  journal={Nature biotechnology},
  volume={42},
  number={3},
  pages={448--457},
  year={2024},
  publisher={Nature Publishing Group US New York}
}

@article{van2024active,
  title={Active inference and deep generative modeling for cognitive ultrasound},
  author={Van Sloun, Ruud JG},
  journal={IEEE Transactions on Ultrasonics, Ferroelectrics, and Frequency Control},
  year={2024},
  publisher={IEEE}
}

@article{duffy2022high,
  title={High-throughput precision phenotyping of left ventricular hypertrophy with cardiovascular deep learning},
  author={Duffy, Grant and Cheng, Paul P and Yuan, Neal and He, Bryan and Kwan, Alan C and Shun-Shin, Matthew J and Alexander, Kevin M and Ebinger, Joseph and Lungren, Matthew P and Rader, Florian and others},
  journal={JAMA cardiology},
  volume={7},
  number={4},
  pages={386--395},
  year={2022},
  publisher={American Medical Association}
}

@article{mamistvalov2022deep,
  title={Deep-learning based adaptive ultrasound imaging from sub-nyquist channel data},
  author={Mamistvalov, Alon and Amar, Ariel and Kessler, Naama and Eldar, Yonina C},
  journal={IEEE Transactions on Ultrasonics, Ferroelectrics, and Frequency Control},
  volume={69},
  number={5},
  pages={1638--1648},
  year={2022},
  publisher={IEEE}
}

@article{huijben2020learning,
  title={Learning sub-sampling and signal recovery with applications in ultrasound imaging},
  author={Huijben, Iris AM and Veeling, Bastiaan S and Janse, Kees and Mischi, Massimo and van Sloun, Ruud JG},
  journal={IEEE Transactions on Medical Imaging},
  volume={39},
  number={12},
  pages={3955--3966},
  year={2020},
  publisher={IEEE}
}

@article{federici2024active,
  title={Active inference for closed-loop transmit beamsteering in fetal Doppler ultrasound},
  author={Federici, Beatrice and van Sloun, Ruud JG and Mischi, Massimo},
  journal={arXiv preprint arXiv:2410.04869},
  year={2024}
}

@article{yousufi2019application,
  title={Application of compressive sensing to ultrasound images: a review},
  author={Yousufi, Musyyab and Amir, Muhammad and Javed, Umer and Tayyib, Muhammad and Abdullah, Suheel and Ullah, Hayat and Qureshi, Ijaz Mansoor and Alimgeer, Khurram Saleem and Akram, Muhammad Waseem and Khan, Khan Bahadar},
  journal={BioMed research international},
  volume={2019},
  number={1},
  pages={7861651},
  year={2019},
  publisher={Wiley Online Library}
}

@article{khan2020adaptive,
  title={Adaptive and compressive beamforming using deep learning for medical ultrasound},
  author={Khan, Shujaat and Huh, Jaeyoung and Ye, Jong Chul},
  journal={IEEE transactions on ultrasonics, ferroelectrics, and frequency control},
  volume={67},
  number={8},
  pages={1558--1572},
  year={2020},
  publisher={IEEE}
}

@inproceedings{ChungKMKY23,
  author       = {Hyungjin Chung and
                  Jeongsol Kim and
                  Michael Thompson McCann and
                  Marc Louis Klasky and
                  Jong Chul Ye},
  title        = {Diffusion Posterior Sampling for General Noisy Inverse Problems},
  booktitle    = {The Eleventh International Conference on Learning Representations,
                  {ICLR} 2023, Kigali, Rwanda, May 1-5, 2023},
  publisher    = {OpenReview.net},
  year         = {2023},
  url          = {https://openreview.net/forum?id=OnD9zGAGT0k},
  bibsource    = {dblp computer science bibliography, https://dblp.org}
}

@inproceedings{stevens2025sequential,
  title={Sequential posterior sampling with diffusion models},
  author={Stevens, Tristan SW and Nolan, Ois{\'\i}n and Robert, Jean-Luc and Van Sloun, Ruud JG},
  booktitle={ICASSP 2025-2025 IEEE International Conference on Acoustics, Speech and Signal Processing (ICASSP)},
  pages={1--5},
  year={2025},
  organization={IEEE}
}

@software{zea2025,
    author = {Stevens, Tristan S.W. and van Nierop, Wessel L. and Luijten, Ben and van de Schaft, Vincent and Nolan, Oisín I. and Federici, Beatrice and van Harten, Louis D. and Penninga, Simon W. and Schueler, Noortje I.P. and van Sloun, Ruud J.G.},
    license = {Apache-2.0},
    month = jul,
    title = {{zea: A Toolbox for Cognitive Ultrasound Imaging}},
    url = {https://github.com/tue-bmd/zea},
    version = {0.0.1},
    year = {2025}
}

@article{ho2020denoising,
  title={Denoising diffusion probabilistic models},
  author={Ho, Jonathan and Jain, Ajay and Abbeel, Pieter},
  journal={Advances in neural information processing systems},
  volume={33},
  pages={6840--6851},
  year={2020}
}

\end{document}